\documentclass[aps,prc,showpacs,showkeys,nofootinbib]{revtex4}
\usepackage{graphicx}% Include figure files
\usepackage{epstopdf}
\usepackage{dcolumn}% Align table columns on decimal point
\usepackage{amsthm}

\usepackage{float}
\usepackage{placeins}
\usepackage{subfigure}
\usepackage{bm}% bold math

\usepackage{amssymb}
\usepackage{amsmath}

\begin{document}

\title{Application of resonant decay method for compound-systems at analysis inclusive spectra in high-energy nuclear reactions}


\author{S.~A.~Omelchenko$^{(1)}$}%
\email{sergomel3000@gmail.com}%
\author{V.~S.~Olkhovsky$^{(1)}$}%
\email{vladislav.olkhovskyy@gmail.com} %
\affiliation{$(1)$Institute for Nuclear Research, National Academy of Sciences of Ukraine, Kyiv, 03680, Ukraine}

\date{\small\today}
%-----------------------------------------------------------------------------------------------------------------------

\begin{abstract}

It is shown, that the exponential decrease of the energy spectra of the reaction final fragments with increasing energy, which does not depend from the fragment's type, targets, projectiles and projectile energies, and which sometimes accompanied slight oscillations, can be explained by the unique phenomenon of time resonances. These time resonances correspond to decay of intermediate excited nuclear composite system.
For the first time, expressions for the decay rate and the probability of survival of such system as a function on time are obtained. Inclusive spectra for protons \textit{p} and helium isotopes $^{3}{\rm He}$ in high-energy nuclear reactions
$^{20}{\rm Ne}$ +$^{238}{\rm U}$ $\rightarrow$ \textit{p} + \textit{X} (2.1 GeV/nucleon) and $^{20}{\rm Ne}$ + $^{238}{\rm U}$ $\rightarrow$ $^{3}{\rm He}$ + \textit{X'} (2.1 GeV/nucleon) are calculated.

\end{abstract}



\maketitle

\section{Introduction
\label{sec.introduction}}

Previously mentioned~\cite{1,2,3,4,5,6,7,8} that for not too heavy particles (from \textit{p} to $^{40}{\rm Ar}$) bombarding targets with energies 0.1-10 GeV/nucleon, the structureless and exponentially falling energy inclusive spectra (sometimes accompanied by light oscillations) of the reaction final fragments were observed. In addition, it was noted that this anomalous phenomenon of high-temperature statistical equilibrium does not depend on the energies of the incident particles either from targets or from the type of fragments of collisional nuclear processes. The same phenomenon was observed for heavier bombarding particles even at lower energies (see, for example,~\cite{9}).

The task of the work is the generalization, development, and further application of the method of time resonances, developed earlier in~\cite{10,11,12} to explain this anomalous behavior of the energy spectra. The basis of the method is the presence of a phenomenological connection between the complex eigenvalues for the metastable states of the time operator $\hat{T}$ and the eigenvalues of the Hamiltonian $\hat{H}$. This connection can be seen from the very form of the Schr\"{o}dinger equation $\hat{H}\Psi =\hat{T}\Psi$. Therefore, the model technique is built on the correspondence of energy resonances with the usual Breit-Wigner form for the amplitude of the reaction (see section \ref{sec.2.1}) to the so-called ``time resonances'' with another parameterization for the amplitude of the reaction using time parameters $t_{n}$ and $\tau_{n}$ with time dimension (see section \ref{sec.2.2}). The meaning of these parameters becomes more apparent from the following general table of correspondences energy resonance and time resonance:

\[\left\{\begin{array}{l} {\hat{H}~~~\leftrightarrow~~\hat{T}} \\ {E~~~\leftrightarrow~~t} \\ {E_{res} \leftrightarrow~~ t_{n} } \\ {{\it \Gamma~~~}\leftrightarrow~~\tau _{n}} \end{array}\right\},\]

where $ E $ - system energy, $t$ - time, $E_{res}$ - energy of the energy resonance, $t_{n}$ - time moment of the time resonance, $\it \Gamma$ - width of the energy resonance and $\tau_{n}$ - width of the time resonance.

The theoretical part of the justification of the method is briefly presented in section \ref{sec.2}. At first we discuss an expression for the amplitude of the reaction for an isolated Breit-Wigner resonance in section \ref{sec.2.1} (see (1) below), which leads to the resonant form of the inclusive spectrum for final fragments of  reaction and to the exponential  dependence on time of decay rate of the intermediate compound system. Then, in  section \ref{sec.2.2}, the amplitude of the reaction of an exponential form ((see \eqref{4} and \eqref{9} below)) is considered, which leads to an exponential type (sometimes with oscillations) of the inclusive spectrum. Besides, it is shown that such parameterization for the reaction amplitude corresponds to the resonant decay on time of an intermediate excited composite system.

The results of calculations of the inclusive spectra of protons \textit{p} and $^{3}{\rm He}$ isotopes emitted in high-energy nuclear reactions in the collision of neon nuclei $^{20}{\rm Ne}$ with uranium nuclei $^{238}{\rm U}$, and their brief analysis are given in section \ref{sec.3}.

In  section \ref{sec.4}, general conclusions are considered, the limits of applicability of the resonant decay model and possible prospects of its development as promising method for analyzing experimental data are discussed.

\section{Resonant and exponential behavior of the amplitude of the reaction in collisional nuclear processes \label{sec.2}}

\subsection{Isolated energy resonance in cross section and the corresponding exponential decay in time of the compound nucleus \label{sec.2.1}}

Before introducing the concept of time resonances, at first we recall that an isolated Breit-Wigner resonance in the cross section of a quantum collision or nuclear reaction \textit{$\alpha$}$\rightarrow$\textit{$\beta$} is related to the exponential decay law on time of the corresponding resonant state of the composite system (see, for example,~\cite{10,13}). In this case, the amplitude of the reaction $f_{\alpha \beta }(E)$ has the form

\begin{equation} \label{1}
\textit{$f_{\alpha \beta } (E)=\frac{C'_{\alpha \beta } }{E-E_{r} +i{\it \Gamma }/2} $},
\end{equation}

where $E_{r} $\textit{ } and \textit{ ${\it \Gamma }$ }-\textit{ }energy and width of resonance; $C'_{\alpha \beta } $ - smooth function of energy \textit{E} in the region ($E_{r} -{\it \Gamma }/2$, $E_{r} +{\it \Gamma }/2$), which, like the reaction amplitude, generally speaking, depends on the emission angle \textit{$\theta$} of the final reaction fragment. For simplicity, wherever this does not lead to ambiguity, the dependence on \textit{$\theta$ }in the arguments of all expressions for the amplitudes and cross sections is omitted. Here energy $E=E_{a} +S_{aC} $ is the excitation energy of the composite system $C=a+A$, consisting of an incident nucleus \textit{a} and a target nucleus \textit{A}, where $E_{a} $ - energy of relative motion of a particle in the input channel, $S_{aC} $ - the energy of separation of the incident particle from the composite system. For decaying compound nucleus the value of \textit{E} determines the energy $E_{b}=E-S_{bC} $ particles \textit{b} emitted into the output channel $b+B=C$, where \textit{B} - compound nucleus residue, $S_{bC} $ - the energy of separation of the ejected particle \textit{b} from the compound nucleus. Therefore, depending on the need, different energies were used as an argument in the amplitude of the reactions, namely, $f_{\alpha \beta } (E)=f_{\alpha \beta } (E_{a} )=f_{\alpha \beta } (E_{b} )$.

In situations with amplitude reactions of the form \eqref{1}, the decay rate of the resonant state (i.e., the probability of decay per unit time) decreases in the usual exponential way on time:

\begin{equation} \label{2} I(t)=({\it \Gamma }/\hbar ){\rm exp}(- {\it \Gamma }t/\hbar ) \end{equation}

under condition

\[{\it \Gamma }<<\Delta E<<E_{r} \; ,\]

where $\Delta$\textit{E} represents the scatter of the energy of incident particles. Differential cross section \textit{$\sigma _{\alpha \beta } $} of the process \textit{$\alpha$}$\rightarrow$\textit{$\beta$} with amplitude \eqref{1} has the form of a Breit-Wigner curve:

\begin{equation} \label{3} \sigma _{\alpha \beta } =\left|f_{\alpha \beta } (E)\right|^{{\rm 2}} =\frac{|C'_{\alpha \beta } |^{2} }{(E-E_{r} )^{2} +{\it \Gamma }^{2} /4} .                                        \end{equation}

\subsection{Exponential decreasing of the reaction amplitude and non-exponential (resonant) decay in time of the composite system
\label{sec.2.2}}

In~\cite{10,11,12}, the probability of the decay of composite systems in processes that was received by such exponentially damped amplitude of the reactions

\begin{equation} \label{4} \; f_{\alpha \beta } (E_{b} )=C_{\alpha \beta } \, {\rm exp}(-E_{b} \tau _{n} /{\rm 2}\hbar +{\rm \; }iE_{b} t_{n} /\hbar )\equiv C_{\alpha \beta } \, {\rm exp}(iE_{b} [t_{n} +i\tau _{n} /{\rm 2}{\rm ]/}\hbar ), \end{equation}

here \textit{$\tau _{n} >0$} and \textit{$t_{n} >0$} - parameters with time dimension, the first of which determines the slope of the exponential dependence of the corresponding differential cross section

\begin{equation} \label{5} \sigma _{\alpha \beta } =\left|f_{\alpha \beta } (E_{b} )\right|^{{\rm 2}} =|C_{\alpha \beta } |^{2} {\rm exp}(-E_{b} \tau _{{\rm n}} /\hbar ).          \end{equation}

Here, as before $C_{\alpha \beta } '$ â \eqref{1}, $C_{\alpha \beta } $ - smooth function of energy, which depend on the emission angle of the particle.

It was shown in~\cite{10,11,12}, that under condition

\begin{equation} \label{6} \tau _{n} <<{\rm 2}\hbar {\rm \; }/\Delta E, \end{equation}

expression \eqref{4} for the amplitude of the reaction corresponds to the resonant decay in time \textit{$t$} of an intermediate excited composite system with a decay rate as a Breit-Wigner function

\begin{equation} \label{7}
I(t)=(2\pi )^{-1} \frac{\tau _{n} }{(t-t_{n} )^{2} +\tau _{n} ^{2} /4}.                                                \end{equation}

The probability of survival of a compound nucleus at time \textit{t} after its formation at a time point $t_{0}$ is described by the following function:

\begin{equation} \label{8}
L_{^{c} } (t)=1-\int _{t_{0} }^{t}dtI(t).
\end{equation}

Therefore, taking into account \eqref{7}, we find

$$
L_{^{c} } (t){\rm \; =\; }1-\pi ^{-1} \left[{\rm arctan}(y)\right]_{y=2t_{0} /\tau _{n} }^{y=2(t-t_{n} -t_{0} )/\tau _{n} } \eqno (8a)
$$

or

$$
L_{^{c} } (t)=1-\pi ^{-1} \left[{\rm arctan}(2(t-t_{n} -t_{0} )/\tau _{n} )+\pi /2\right] \eqno (8b)                             $$

and

$$
L_{^{c} } (t)=\left\{\begin{array}{l} {1,\begin{array}{ccc} {} & {} & {} \end{array}{\rm \; when}\mathop{}\limits_{} \mathop{}\limits_{} 0\le t<t_{n} \mathop{}\nolimits^{} ({\rm at}\mathop{}\limits_{} -2t_{0} /\tau _{n} \mathop{}\limits_{} \to \infty )\mathop{}\limits_{} {\rm and}^{} } \\ {0,\begin{array}{ccc} {} & {} & {} \end{array}{\rm \; when}\mathop{}\limits_{} \mathop{}\limits_{} t\to \infty .} \end{array}\right.
\eqno (8c)
$$

As can be seen from \eqref{7} and the results presented below in Fig.~\ref{fig.1}, the decay rate of the intermediate system, which is formed in nuclear processes with an exponentially decreasing amplitude of the reaction \eqref{4}, is resonant in nature. An intermediate nuclear complex consisting, generally speaking, of fragments of a bombarding particle and fragments of a target nucleus that forms after a collision at time $t_{0}$ and exists, on average, during time $t_{n} -t_{0} $, decays during the time comparable to the width $\tau_{n}$ of the decay rate $I(t)$.

By analogy with the energy resonance considered in the previous section 2.1, the phenomenon of formation of intermediate systems with a resonant decay rate as a Breit-Wigner function \eqref{7} is called a ``time resonance'' with parameters $t_{n}$ and $\tau_{n}$, which correspond to its average decay time and uncertainty in the time of this decay time. Note that when the width of the time resonance tends to zero ($\tau _{n} \to 0$), the decay rate changes to a delta function, which corresponds to the instantaneous (explosive) nature of the decay of the intermediate compound-complex. This led to the use in our previous works~\cite{10,11,12} of the term ``time explosions'' as an equivalent of the term ``time resonances''.

\begin{figure}[htbp]
\centerline{\includegraphics[width=85mm]{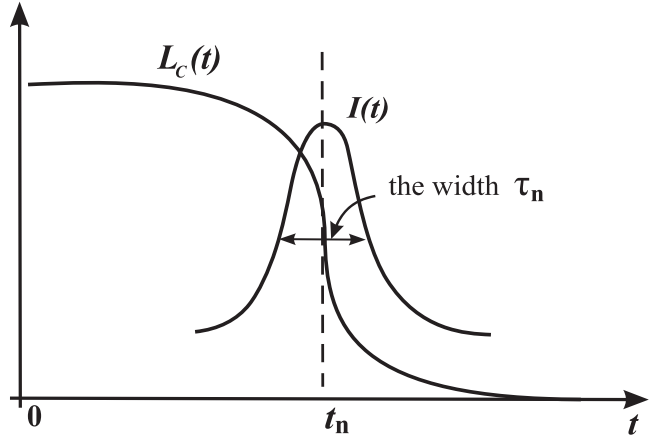}}
\caption {Schematic representation for dependences on time of the decay rate
  $I(t)$ and survival probability $L_{c}(t)$ for intermediate system when $t_{0}=0$.
\label{fig.1}}
\end{figure}

Note that for simplicity in Fig.~\ref{fig.1} (see also Fig.~\ref{fig.3} below), the value $t_{0} $ is assumed to be zero for correct combination of images with arguments actually shifted by value $t_{0} $ in functions $L_{c} (t-t_{0})$ and $I(t)$ (see \eqref{8}).

As noted in the introduction, inclusive high-energy spectra can oscillate. Such oscillations can be described using a multi-component expression for the amplitude of the reaction of the form, representing a more complex linear combination of one-component expressions of the form \eqref{4} with $\nu >1$:

\begin{equation} \label{9}
f_{\alpha \beta } (E_{b} )=\sum _{n=1}^{\nu }C_{\alpha \beta } ^{n} \exp [-E_{b} \tau _{n} /2\hbar +iE_{b} t_{n} /\hbar].
\end{equation}

In such a situation, the differential cross section \textit{$\sigma _{\alpha \beta } =\left|f_{\alpha \beta } \right|^{{\rm 2}} $ }contains not only exponentially decreasing terms, but also oscillating terms with multipliers ${\rm cos}[E_{b} (t_{n} - t_{n'} )/\hbar ]$ or \textit{${\rm sin}[E_{b} (t_{n} - t_{n'} )/\hbar ]$},
and, for example, in the case of two terms in \eqref{9} for $\nu =2$, the formula for the section will take the following form:

\begin{equation} \label{10}
\begin{array}{lcl}
\sigma _{\alpha \beta } =\left|C_{\alpha \beta } ^{1} \right|^{{\rm 2}} \; {\rm exp}(-E_{b} \tau _{1} /\hbar )+\left|C_{\alpha \beta } ^{2} \right|^{{\rm 2}} \; {\rm exp}(-E_{b} \tau _{2} /\hbar )\; +\; \\ \; \\
\; \; +\;
{\rm 2Re} \{
C_{\alpha \beta }^{1}C_{\alpha \beta }^{2*} {\rm exp}[(i(t_{{\rm 1}} - {\rm \; }t_{{\rm 2}} )-{\rm \; }(\tau _{{\rm 1}} +\tau _{{\rm 2}} )/{\rm 2)}\, E_{b} /\hbar ]
\}.

\end{array}
\end{equation}

For the multicomponent amplitude \eqref{9}, we write the following expression for the decay rate of the intermediate composite system for the case of several time resonances without taking into account their interference (see \eqref{eq.app.2.1.10} in the Appendix~\ref{sec.app.1} below):

\begin{equation} \label{11}
I(t)=\left(2\pi \sum _{n=1}^{\nu }\frac{\left|C_{\alpha \beta }^{n} \right|^{2} }{\tau _{n} }  \right)^{-1} \sum _{n=1}^{\nu }\frac{\left|C_{\alpha \beta }^{n} \right|^{2} }{(t-t_{n} )^{2} +\tau _{n} ^{2} /4}.
\end{equation}

From where, taking into account (8b), we can write the expression for the probability of survival $L_{c} (t)$ in the following form:

\begin{equation} \label{12}
L_{c} (t)=1-\left(\pi \sum _{n=1}^{\nu }\frac{\left|C_{\alpha \beta }^{n} \right|^{2} }{\tau _{n} }  \right)^{-1} \sum _{n=1}^{\nu }\left[\frac{\left|C_{\alpha \beta }^{n} \right|^{2} }{\tau _{n} } \left[{\rm arctan}(2(t-t_{n} -t_{0} )/\tau _{n} )+\pi /2\right]\right].
\end{equation}

\section{Calculations of the inclusive energy spectra of protons and helium isotopes for two high-energy reactions
\label{sec.3}}

It should be noted, for a complete analysis of the inclusive experimental spectra of reaction fragments, it is necessary to take into account the possible dependence of the amplitude of the reaction \eqref{9} on various quantum numbers (for example, total angular momentum, spins, orbital angular momentum, etc.) and also take into account all possible intermediate processes for every particular nuclear reaction with emission of the observed final fragment.

In this section, we analyze the inclusive spectra of protons and helium isotopes of high-energy nuclear reactions $^{20}{\rm Ne}$+$^{238}{\rm U}$ $\rightarrow$ \textit{p}+\textit{X} (2.1 GeV/nucleon) and $^{20}{\rm Ne}$+ $^{238}{\rm U}$ $\rightarrow$ $^{3}{\rm He}$+\textit{X'} (2.1 GeV/nucleon). These spectra decrease exponentially with energy and oscillate slightly. Therefore, to describe them, we will use the expression \eqref{10}.

\begin{figure}[htbp]
\centerline{\includegraphics[width=85mm]{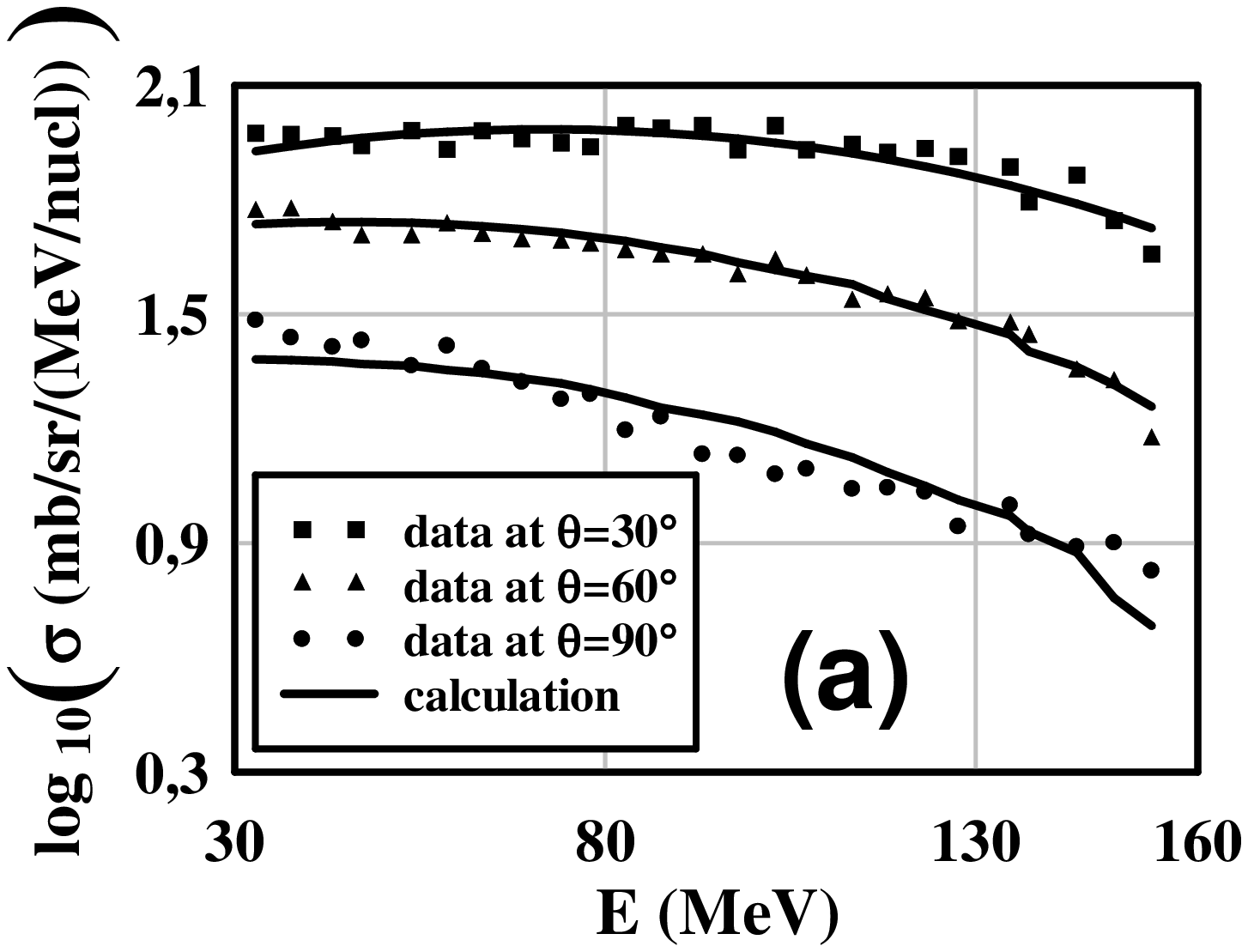}
\hspace{0mm}\includegraphics[width=85mm]{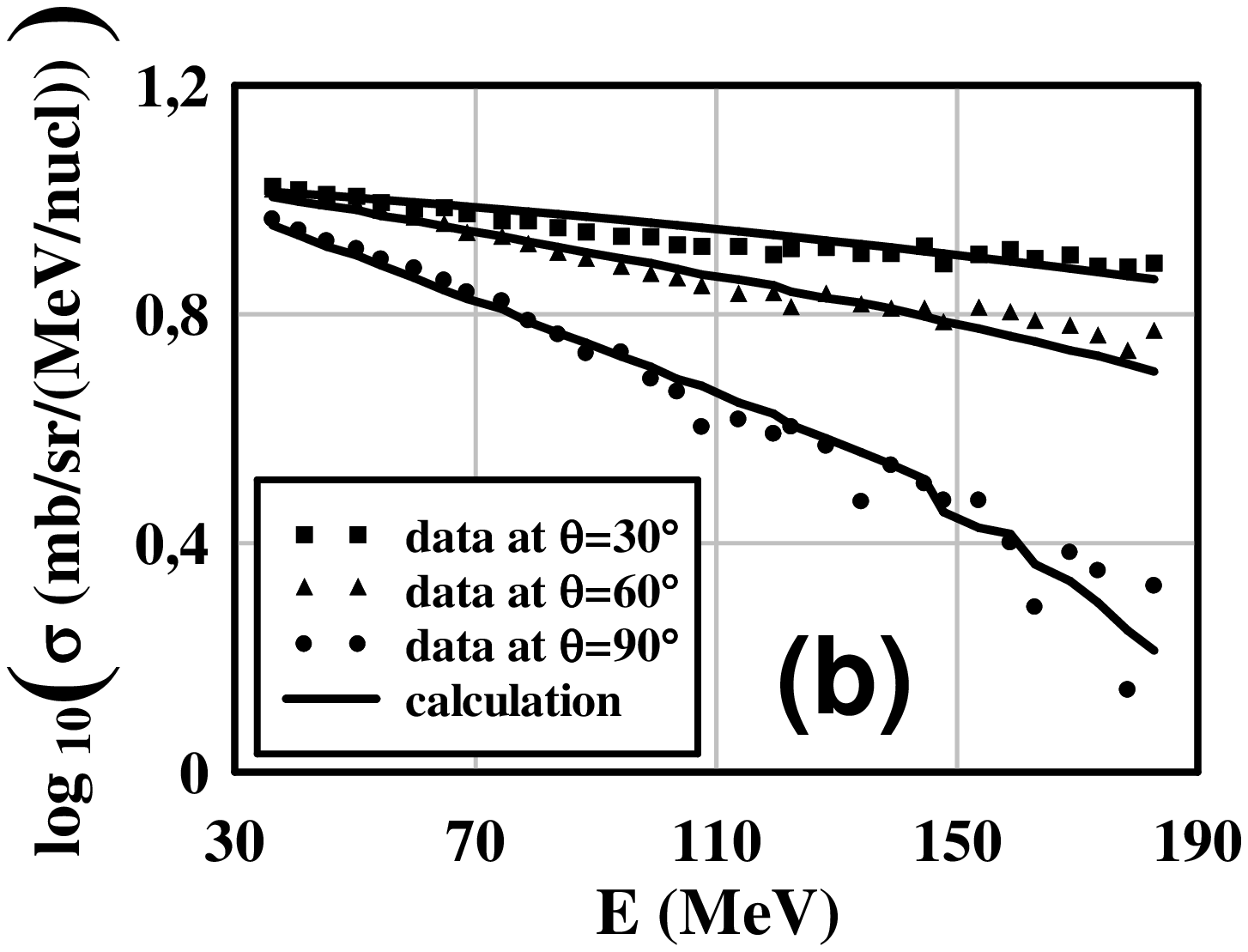}}
\vspace{-2mm}
\caption{Inclusive spectra of different final fragments for collision process $^{20}{\rm Ne}$ +$^{238}{\rm U}$, namely: [Panel a] - inclusive energy spectrum of proton and
[Panel b] - inclusive energy spectrum of helium isotope $^{3}{\rm He}$.
\label{fig.2}}
\end{figure}

Fig.~\ref{fig.2} show some calculated inclusive energy spectra \textit{$\sigma _{\alpha \beta } $} in comparison with experimental data taken from~\cite{14}.

 Designation \textit{$\theta$ }in Fig.~\ref{fig.2} is the departure angle of the final fragment in the laboratory system. The values of the parameters in \eqref{10} were found from fitting the theoretical curves to the experimental data and are listed in Tables~\ref{tabular:table1} and~\ref{tabular:table2}. The values of the time parameters $\tau_{1}$, $\tau_{2}$ and
$t_{2} -t_{1} $ are in units $10^{-23}$ sec.

\renewcommand{\arraystretch}{1.5}

\begin{table} [h]

\begin{minipage}[t]{.39\textwidth}
%\parindent=2em
%\hspace{-2pt}
\vspace{0pt}
\caption{Calculation parameters for inclusive process with emission of protons.}
\label{tabular:table1}
\begin{tabular}{c c c c c c}
\hline \hline
\textit{$\theta $} & \textit{$\tau _{1} $} & \textit{$\tau _{2} $} & \textit{$t_{{\rm 2}} -t_{{\rm 1}} $} & $C_{\alpha \beta }^{1} $ & $C_{\alpha \beta }^{2} $  \\ \hline
$~30^{\circ}~~$\newline  &  ~~0.1~~  &  ~~0.1~~  &  ~1.2~  &~(5.2,0.4) ~& (0.1,5.2)  \\
$60^{\circ}$\newline  & 0.1 & 0.1 & 1.2 & (3.7,1.1) & (1.1,3.7)    \\
$90^{\circ}$\newline  & 0.1 & 0.1 & 1.2 & (2.6,1.5) & (1.5,2.6)    \\
\hline  \hline
\end{tabular}
\end{minipage}
\hspace{17mm}
%\hfill
\begin{minipage}[t]{.39\textwidth}
\vspace{0pt}
\caption{Calculation parameters for inclusive process with emission of helium isotopes.}
\label{tabular:table2}
\begin{tabular}{c c c c c c}
\hline \hline
\textit{$\theta $} & \textit{$\tau _{1} $} & \textit{$\tau _{2} $} & \textit{$t_{{\rm 2}} -t_{{\rm 1}} $} & $C_{\alpha \beta }^{1} $ & $C_{\alpha \beta }^{2} $  \\ \hline
$~30^{\circ}~~$ \newline  &  ~~0.1~~  &  ~~0.2~~  &  ~0.3~  &~(1.5,0.6)~ & (1.2,1.3) \\
$60^{\circ}$\newline  & 0.1 & 0.2 & 0.3 & (0.9,1.5) & (1.5,0.9) \\
$90^{\circ}$\newline  & 0.1 & 0.2 & 0.3 & (0,2.45) & (2.45,0) \\
\hline  \hline
\end{tabular}
\end{minipage}

\end{table}

\begin{figure}[htbp]
\centerline{\includegraphics[width=85mm]{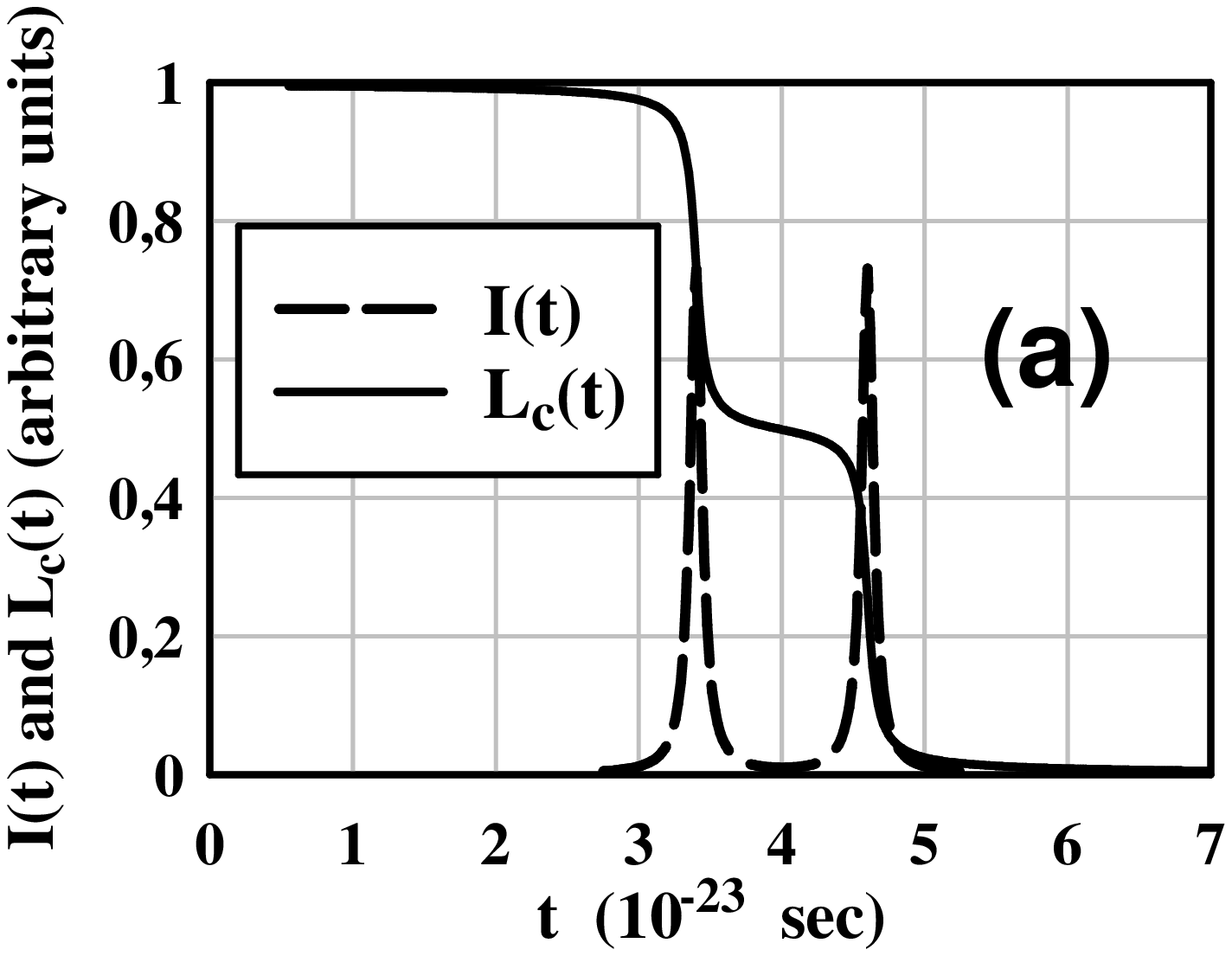}
\hspace{2mm}\includegraphics[width=85mm]{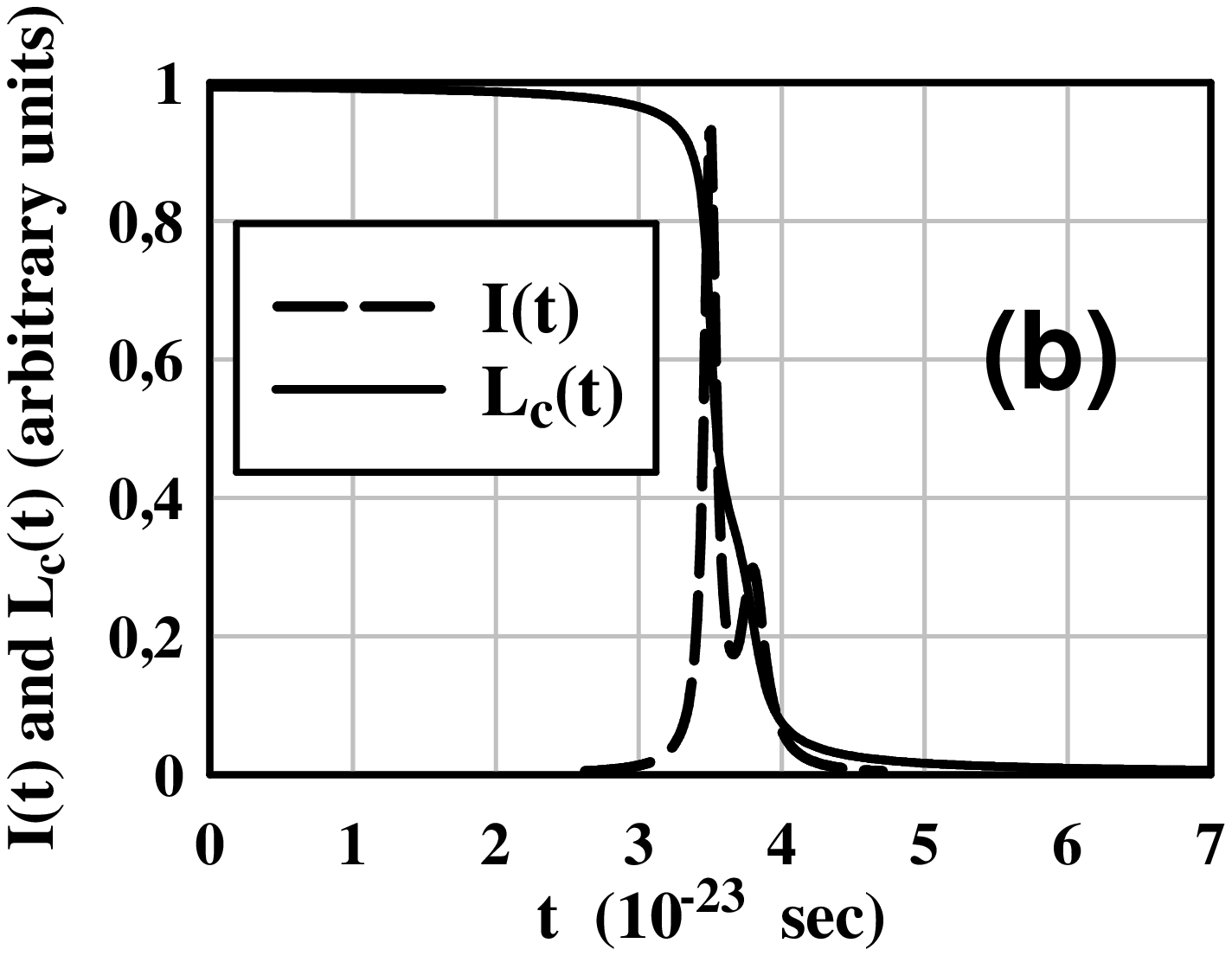}}
\vspace{-2mm}
\caption{Decay rate $I(t)$ and survival probability $L_{c}(t)$ for
two different inclusive nuclear processes, namely:
[Panel a] - for reaction
$^{20}{\rm Ne}$ + $^{238}{\rm U}$ $\rightarrow$ \textit{p}+\textit{X}
(at $t_{1} =3.4\cdot 10^{-23} $sec, $t_{0} =0$, $\theta =30^{{\bf o}} $);
[Panel b] - for reaction
$^{20}{\rm Ne}$ + $^{238}{\rm U}$ $\rightarrow$ $^{3}{\rm He}$ + \textit{X'}
(at $t_{1} =3.5\cdot 10^{-23} $sec, $t_{0} =0$, $\theta =30^{{\bf o}} $).
\label{fig.3}}
\end{figure}

In Fig.~\ref{fig.3} there are graphs of the change in time of the decay rate $I(t)$ \eqref{11} and survival probability $L_{c}(t)$ \eqref{12} for the considered inclusive processes with parameters from Tables~\ref{tabular:table1} and~\ref{tabular:table2}.
Approximately the same parameters $t_{1}$ for both reactions were chosen for reasons of convenience of comparing the results. From Fig.~\ref{fig.3}, we can conclude that the choice of the two-component type \eqref{9} for the amplitude of the reaction \textit{$\; f_{\alpha \beta } (E)$} only then does not fundamentally change the nature of the probability of survival $L_{c} (t)$ (shown previously in Fig.~\ref{fig.1} for the single-component amplitude of the reaction (4)), when closely spaced ``strongly overlapping'' time resonances are selected with values $t_{2} -t_{1} $, close to values $\tau _{1,2} $ (see. Tables~\ref{tabular:table1} and~\ref{tabular:table2}). This conclusion is consistent with a clearer fall off in the probability of survival in the Fig.~\ref{fig.3}~(a) than in the Fig.~\ref{fig.3}~(b).

\section{Discussion, conclusions and perspectives}
\label{sec.4}

1. The phenomenological time approximation developed in this paper is based on a generalization of the results of a joint time and statistical analysis of energy resonances in nuclear reactions. This approximation based on time resonances can be completely combined with any semi-microscopic model based, for example, on a fireball model or on an intranuclear cascade model.

2. In this paper, we presented some new results of using the method of time resonances to calculate real inclusive spectra of the final fragments of scattering reactions at high energies. The results of our calculations are in satisfactory agreement with the experimental data. For the first time, expressions for the decay rate $I(t)$ (see \eqref{11}) and the probability of survival $L_{c} (t)$ (see \eqref{12}) are obtained in the presence of an arbitrary number $\nu >1$ of time resonances.

3. Since the parameters of time resonances ($t_{n}$ and $\tau_{n}$) are internal characteristics of the state for the highly excited intermediate nuclear compound system, then the dependence on the emission angle $\theta$ of the investigated final fragment of the high-energy nuclear reaction presents only in $C_{\alpha \beta }^{n} $. Therefore, the search of asymptotic or analytical dependence of the quantities $C_{\alpha \beta }^{n} $ on $\theta$ is a very interesting direction for the further development of the method, both from a theoretical point of view and from the point of view of minimizing the number of fitting parameters at the study of various new experimental inclusive spectra of the final fragments of high-energy reactions.

4. It is important to note the method presented here has limits of applicability. An experimental indicator of the presence of time resonances is the presence of exponentially decaying energy inclusive spectra of final particles (sometimes with small oscillations in the presence of several time resonances).
In addition, our resonant decay model occupies an important intermediate cross-linking position on the energy scale of scientific research, namely, in a fairly wide energy range of 0.1-10 GeV/nucleon, which is an intermediate boundary zone between of using fireball models with lower energies and nuclear hydrodynamic models with higher ones.

5. At last, the parameters of time resonances ($t_{n}$ and $\tau_{n}$) corresponding, generally speaking, to a particular metastable state of a time operator, are fixed for every particular reaction. In our opinion, the probability of various reactions going through the same set of intermediate states (i.e. the same set of pairs of time parameters ($t_{n}$ and $\tau_{n}$) of the investigated compound system is minimal. This is true even for this work, where two reactions with the same input channels are considered (i.e. $^{20}{\rm Ne}$ +$^{238}{\rm U}$).

\section*{Acknowledgements
%\section{Acknowledgements
\label{sec.acknowledgements}}

We express our gratitude and deep appreciation to Professor V.A. Pluyko (Taras Shevchenko National University, Kyiv, Ukraine) and to Dr. S.P. Maydanyuk (Institute for Nuclear Research, Kyiv, Ukraine) for valuable advice and comments on the article.

\appendix
\section{Decay rate function of an excited composite system for arbitrary number of pairs time resonances $t_{n}$ and $\tau_{n}$
\label{sec.app.1}}

Using the same technique from~\cite{10} when obtaining formula \eqref{7} for the decay rate $I(t)$ of an excited composite system in the case of a one-component amplitude of reaction \eqref{4} with one time resonance, we obtain here expression \eqref{11} (see here \eqref{eq.app.2.1.10}) for $I(t)$ in the case of choosing a multi-component amplitude of reaction \eqref{9} with several time resonances in the approximation of the absence of interference between them.

As already indicated in~\cite{10}, when choosing the one-component amplitude of the reaction with time parametrization of the form \eqref{4}, the wave packet of the final particle in the one-dimensional radial asymptotic limit is as follows (see, for example, (24) in~\cite{10}):

\begin{equation}
\Psi _{\beta } (R_{\beta } ,t)\approx \int _{E_{\min } }^{\infty }dE'g(E')C_{\alpha \beta } \exp [-iE'\tau _{n} /2+iE'(t_{n} -t)/\hbar ],
\label{eq.app.2.1.1}
\end{equation}

where $R_{\beta }$ -- interaction radius in the final channel, $C_{\alpha \beta } $ - constant or very smooth function inside $\Delta E$ from particle energy \textit{E} in the final channel. If the function \textit{g}(\textit{E'}) choose in the simplest rectangular view (see. (25) in~\cite{10}):

\begin{equation}
\begin{array}{l} {g(E')=\left\{\begin{array}{l} {\left(\Delta E\right)^{-1/2} \exp (i\arg \mathop{}\limits^{} g),\begin{array}{cc} {} & {} \end{array}E_{\min } \le E-\Delta E/2<E'<E+\Delta E/2} \\ \; \\ {0,\begin{array}{cccc} {} & {} & {} & {}  \end{array}\begin{array}{cccc} {} & {} & {} & {} \end{array}{\rm \; \; \; \; \; \; \; \; \; \; \; \; \; \; \; \; \; \; \;}E'<E-\Delta E/2{\rm ,\; }E'>E+\Delta E/2}, \end{array}\right . } \\
\end{array}
\label{eq.app.2.1.2}
\end{equation}

where arg\textit{ g} is a smooth function of \textit{E} inside $\Delta$\textit{E}, then we get (see. (26) in ~\cite{10})

\begin{equation} \label{2.1.3}
\begin{array}{lcl}
\; \; \; \; \; \; \; \; \;\Psi_{\beta}(R_{\beta},t)=\displaystyle\frac{C_{\alpha \beta}}{t-t_{n}+i\tau_{n}/2} \exp [E(-\tau_{n}/2+i(t_{n}-t))/\hbar ]\times \; \; \\ \; \\
\times
\left[\exp \left[\Delta E(-\tau _{n} /2+i(t_{n} -t)/2\hbar )-\exp \left[-\Delta E(-\tau _{n} /2+i(t_{n} -t)/2\hbar )\right. \right]\right].
\end{array}
\end{equation}

If all energies are in a large interval, starting from \textit{E}min, completely filled, i.e.

\begin{equation}
\begin{array}{lcl}
\left\{\begin{array}{l}
{(E+\Delta E/2)\tau_{n}/2\mathop{}\limits^{} \to \infty \mathop{}\limits^{} } \\ {E-\Delta E/2\begin{array}{ccc} {} & {} & {} \end{array}\to E_{\min } } \end{array}\right.,
\end{array}
\label{eq.app.2.1.4}
\end{equation}

then we get for the case of a single time resonance (see., for example, (28) in~\cite{10})\textit{}

\begin{equation}
\Psi_{\beta}(R_{\beta},t)=\frac{C_{\alpha \beta}}{t-t_{n}+i\tau_{n}/2} \exp [E_{\min}(-\tau_{n}/2+i(t_{n}
-t))/\hbar].
\label{eq.app.2.1.5}
\end{equation}

For the case of several time resonances with the amplitude of reaction \eqref{9}, instead of (A1), we can write

\begin{equation}
\Psi _{\beta } (R_{\beta } ,t)\approx \int _{E_{\min } }^{\infty }dE'g(E')\left[\sum _{n=1}^{\nu }C_{\alpha \beta }^{n} \exp [-iE'\tau _{n} /2+iE'(t_{n} -t)/\hbar ] \right]
\label{eq.app.2.1.6}
\end{equation}

where, in accordance with the above calculations (\eqref{eq.app.2.1.1}-\eqref{eq.app.2.1.5}), we obtain the expression

\begin{equation}
\Psi _{\beta } (R_{\beta } ,t)=\sum _{n=1}^{\nu }\frac{C_{\alpha \beta }^{n} }{t-t_{n} +i\tau _{n} /2} \exp [E_{\min } (-\tau _{n} /2+i(t_{n} -t))/\hbar ].
\label{eq.app.2.1.7}
\end{equation}

Then, taking into account the expression for the flux density of scattered particles (see, for example, (37) in ~\cite{10}),

\begin{equation}
%\begin{array}{lcl}
j_{\beta } (R_{\beta } ,t)=Re\left[\Psi _{\beta } (R_{\beta } ,t)\frac{i\hbar }{m_{\beta } } \mathop{\lim }\limits_{z_{\beta } \to R_{\beta } } \frac{\partial \Psi _{\beta }^{\bullet } }{\partial z_{\beta } } \right]\cong \bar{\upsilon }\left|\Psi _{\beta } (R_{\beta } ,t)\right|^{2}
%\end{array}
\label{eq.app.2.1.8}
\end{equation}

(here, $z_{\beta } $- the axis along the direction of motion of the final particle in accordance with the chosen registration geometry, $\bar{\upsilon }$ is the average propagation velocity of the wave packet of the final particle) and, using the general expression for the decay rate of the composite system in its vicinity (about $z_{\beta } \approx R_{\beta } $, see, for example, (39) in~\cite{10}), we write

\begin{equation}
%\begin{array}{lcl}
I\left(t\right){\rm \; }=\frac{j_{\beta } (R_{\beta } ,t)}{\int _{-\infty }^{\infty }dtj_{\beta } (R_{\beta } ,t) } =\frac{\left|\Psi _{\beta } (R_{\beta } ,t)\right|^{2} }{\int _{-\infty }^{\infty }dt\left|\Psi _{\beta } (R_{\beta } ,t)\right|^{2}  } \approx \frac{\sum _{n=1}^{\nu }\left|C_{\alpha \beta }^{n} \right|^{2} ((t-t_{n} )^{2} +\tau _{n} ^{2} /4)^{-1} +\Delta }{\int _{-\infty }^{\infty }dt\left[\sum _{n=1}^{\nu }\left|C_{\alpha \beta }^{n} \right|^{2} ((t-t_{n} )^{2} +\tau _{n} ^{2} /4)^{-1} +\Delta \right] },
%\end{array},
\label{eq.app.2.1.9}
\end{equation}

whence, in the approximation of neglecting interference terms $\Delta$, we get

\begin{equation}
I\left(t\right){\rm \; }\approx \frac{\sum _{n=1}^{\nu }\left|C_{\alpha \beta }^{n} \right|^{2} ((t-t_{n} )^{2} +\tau _{n}^{2} /4)^{-1}  }{\int _{-\infty }^{\infty }dt\left[\sum _{n=1}^{\nu }\left|C_{\alpha \beta }^{n} \right|^{2} ((t-t_{n} )^{2} +\tau _{n}^{2} /4)^{-1}  \right] } =\left(2\pi \sum _{n=1}^{\nu }\frac{\left|C_{\alpha \beta }^{n} \right|^{2} }{\tau _{n} }  \right)^{-1} \sum _{n=1}^{\nu }\frac{\left|C_{\alpha \beta }^{n} \right|^{2} }{(t-t_{n} )^{2} +\tau _{n} ^{2} /4}.
%\end{array}.
\label{eq.app.2.1.10}
\end{equation}

\end{document}